\documentclass[final]{cvpr}
\usepackage{booktabs} 
\usepackage{siunitx}  
\usepackage{times}
\usepackage{epsfig}
\usepackage{graphicx}
\usepackage{amsmath}
\usepackage{amssymb}
\usepackage{array}
\usepackage{graphicx}
\usepackage{multirow}
\usepackage{caption} 

\newcommand\MyBox[2]{
  \fbox{\lower0.75cm
    \vbox to 1.7cm{\vfil
      \hbox to 1.7cm{\hfil\parbox{1.4cm}{#1\\#2}\hfil}
      \vfil}%
  }%
}
\graphicspath{ {./images/} }

\usepackage[pagebackref=true,breaklinks=true,colorlinks,bookmarks=false]{hyperref}

\def\cvprPaperID{****} 
\def\confYear{CVPR 2021}

\begin{document}

\title{Hybrid(Transformer+CNN)-based Polyp Segmentation}

\author{Madan Baduwal \\
\textsuperscript{\rmfamily\textbf{}}Mississippi State University\\
\texttt{mb4239@msstate.edu} \\
\url{https://madanbaduwal.github.io/polyp-seg} 
}



\makeatletter
\let\@oldmaketitle\@maketitle
\renewcommand{\@maketitle}{
    \@oldmaketitle
    \begin{center}
        \includegraphics[width=\textwidth]{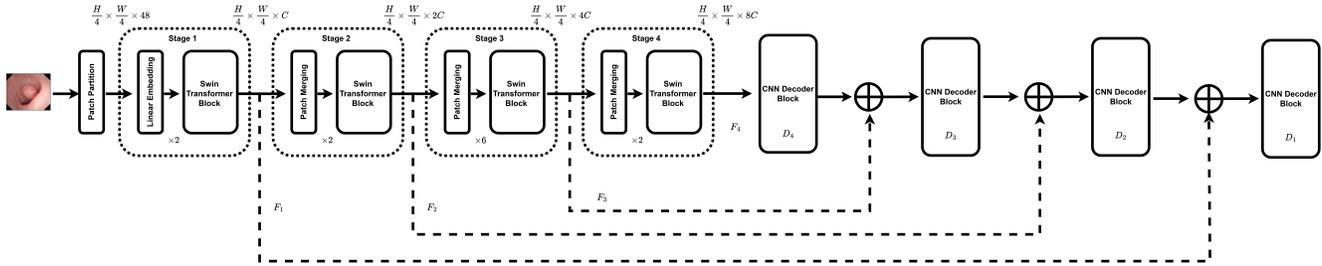}
        \captionof{figure}{Hybrid (Transformer+CNN)-based Polyp Segmentation Architecture}
        \label{fig:wide_image}
    \end{center}
}
\makeatother

\maketitle

\begin{abstract}
Colonoscopy is still the main method of detection and segmentation of colonic polyps, and recent advancements in deep learning networks such as U-Net, ResUNet, Swin-UNet, and PraNet have made outstanding performance in polyp segmentation. Yet, the problem is extremely challenging due to high variation in size, shape, endoscopy types, lighting, imaging protocols, and ill-defined boundaries (fluid, folds) of the polyps, rendering accurate segmentation a challenging and problematic task. To address these critical challenges in polyp segmentation, we introduce a hybrid (Transformer + CNN) model that is crafted to enhance robustness against evolving polyp characteristics. Our hybrid architecture demonstrates superior performance over existing solutions, particularly in addressing two critical challenges: (1) accurate segmentation of polyps with ill-defined margins through boundary-aware attention mechanisms, and (2) robust feature extraction in the presence of common endoscopic artifacts including specular highlights, motion blur, and fluid occlusions. Quantitative evaluations reveal significant improvements in segmentation accuracy (Recall improved by 1.76\%, i.e., 0.9555, accuracy improved by 0.07\%, i.e., 0.9849) and artifact resilience compared to state-of-the-art polyp segmentation methods.
\end{abstract}


\section{Introduction}

Colon polyps are neoplastic polyps developing from the lining epithelium of the colon or rectum. Incidence epidemiological data suggest a prevalence of approximately 30\% among individuals over the age of 50 years ~\cite{1}. Although, in most cases, they are benign, a proportion among them, especially adenomatous polyps, are liable to malignancy by the stepwise adenoma-carcinoma sequence. This shift carries critical clinical import, as CRC is still the third most diagnosed malignancy within the United States, with 38 per 100,000 cases and a commensurate rate of death being 13 per 100,000 annually ~\cite{1}. Colonoscopy is the cornerstone of preventing CRC, with the ability to detect and treat concomitantly via eliminating pre-malignant polyps. Its use in the initial treatment decreases CRC mortality and incidence by quite a lot ~\cite{2}, showcasing its value in practice.

Colonoscopy is a minimally invasive but very effective diagnostic procedure for detecting colorectal polyps when performed by skilled endoscopists. While helpful, existing colonoscopy tests fail in the detection of 22-28 \% of polyps ~\cite{3}, which could develop into advanced cancers and deteriorating clinical outcomes. The process entails using a colonoscope, a finger-thick, flexible tube containing a light source, and a video camera passed transanally to observe the colorectal mucosa. Integrated working channels allow concurrent removal of polyps or biopsy when suspicious lesions are encountered ~\cite{3}.

Polyp shape is highly heterogeneous during stages of development. As seen in Fig. ~\ref{fig:segmentation_results}, variation of structural characteristics, size measurements, and color characteristics provides difficult diagnostic issues at a high level. These are the most demanding challenges for low-contrast small polyps ($<5$ mm) in size. These constraints lead to low rates of detection, even being carried out by trained operators using standard image gathering methods.

The clinical need for accurate segmentation of polyps beyond detection is the need to precisely demarcate lesion boundaries to direct therapeutic interventions. Online segmentation of polyp architectures (e.g., discrimination between adenomatous and hyperplastic growth patterns) would directly influence intraoperative decisions, allowing for the complete excision of neoplastic lesions with sparing of normal tissue when polyps are not neoplastic. Despite convolutional and transformer-based models achieving $>90$\% Dice scores on offline testing, clinical adoption is strictly limited by some inherent limitations. Current architectures cannot maintain diagnostic-grade segmentations in true real-time endoscopy settings primarily because of the computational delay involved in the processing of 1080p video streams at 30fps. This is compounded by morphological nuances: indistinct boundaries on sessile or flat polyps decrease segmentation accuracy by 15–20\% compared with pedunculated lesions, and motion artifacts and specular reflections also degrade boundary precision. This underlying conflict between model complexity (to provide pixel-level precision) and inference speed (to provide clinical usability) identifies the medical imperative for lightweight but robust segmentation algorithms in gastrointestinal endoscopy.

To address these challenges, our work presented in this paper contributes the following:

\begin{itemize}
    \item We suggest a hybrid approach in which pre-trained vision transformers are utilized for global feature extraction and light-weight CNNs for spatial fine-tuning to achieve precise polyp segmentation and minimize computational complexity.
    \item Achieved the highest frames-per-second (FPS) alongside state-of-the-art (SOTA) results in performance metrics on various datasets (such as the Kvasir-SEG dataset ~\cite{4}), compared to other SOTA models like NanoNet ~\cite{5}, ResUNet++ ~\cite{7}, and ResUNet++ + CRF ~\cite{8}.
\end{itemize}

\begin{figure}[htbp]
   \addtocounter{figure}{1} 
   \centering
   \includegraphics[width=0.7\linewidth]{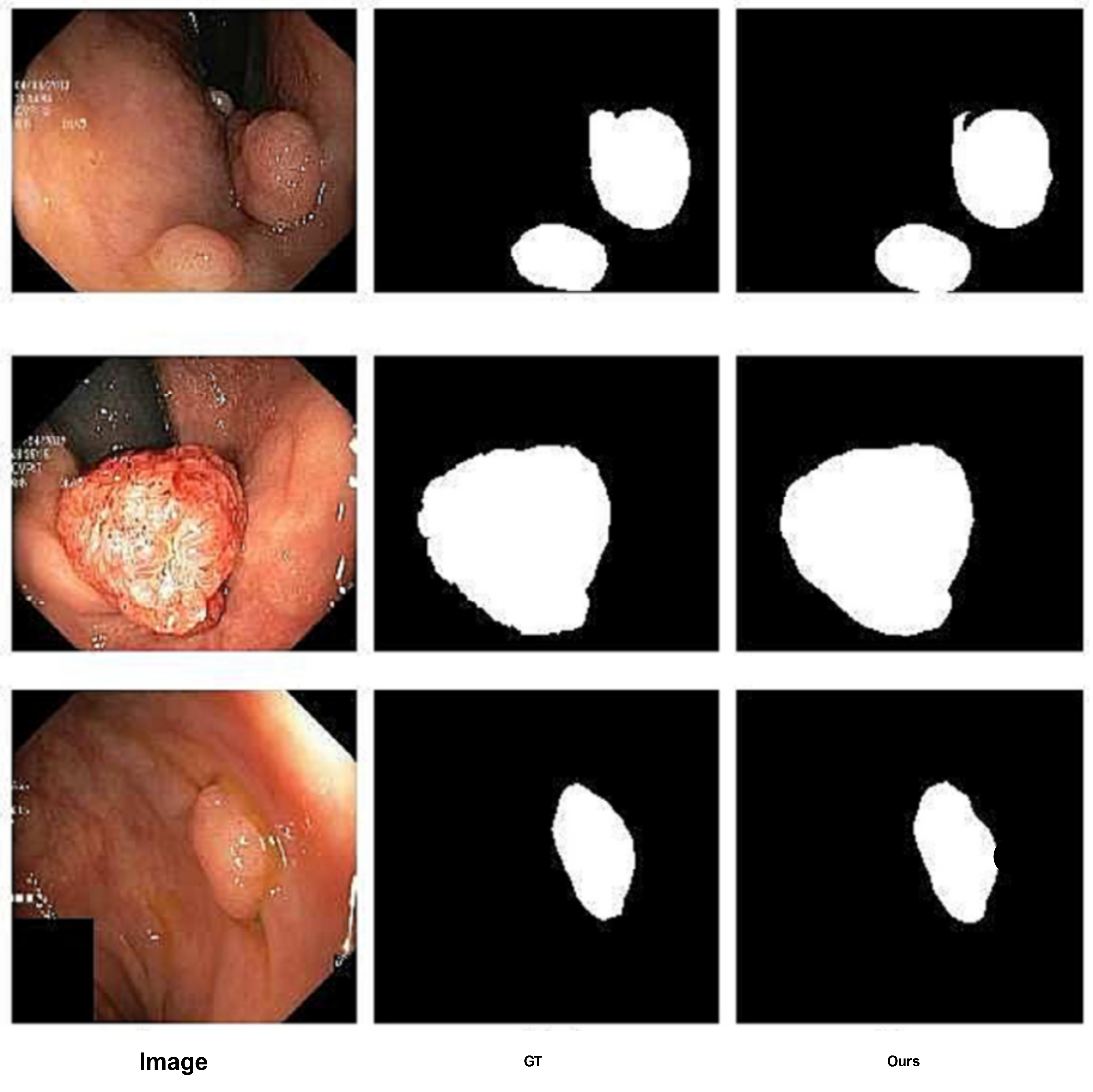} 
   \caption{Comparative visualization of polyp segmentation results on the Kvasir-SEG dataset. From left to right: (a) Original endoscopic images, (b) Pixel-level ground truth annotations, and (c) Predicted segmentation masks generated by our proposed model.}
   \label{fig:segmentation_results}
\end{figure}


\section{Related work}

\subsection{Classic methods}

Early works proposed methods to address the problem of polyp segmentation using classical image processing techniques~\cite{9}. However, these methods struggled to achieve satisfactory performance due to the similarity between the polyps and the surrounding background.

\subsection{Convolution networks}

Deep learning techniques~\cite{10, 11, 12} have immensely enhanced the performance of polyp segmentation. U-Net model~\cite{13} with its encoder-decoder in generic form and skip connections is a smooth benchmark for segmentation of polyps. Its variations, such as U-Net++~\cite{zhou2018unet++}, used nested skip connections to visualize fine details while ResUNet~\cite{zhang2018road} and ResUNet++~\cite{7} made use of residual blocks to achieve smooth gradient propagation and feature discovery. DoubleU-Net~\cite{jha2020doublenet} further enhanced segmentation using a two-stage U-Net model, further advancing the detection of small polyps.

Recent advancements have concerned attention mechanism boost and contextual modeling. PraNet~\cite{fan2020pranet} proposed reverse attention and boundary refining and achieved SOTA performance in a number of polyp benchmarks. DDANet~\cite{liu2021ddanet} also utilized dilated dual attention blocks to boost contextual modeling, and UACANet-S/L~\cite{oh2021uacanet} utilized channel attention to concentrate on polyp regions. For addressing size variation, MSRF-Net and MSRFE-Net~\cite{wang2022msrfe} utilized multi-scale residual fusion and boosted segmentation of various polyp sizes.

Real-time segmentation effort has also been made. Jha et al.~\cite{5} employed Conditional Random Field (CRF) post-processing for enhancing contextual information, whereas Thambawita et al.~\cite{15} proposed pyramid-based augmentation for generalization. ColonSegNet~\cite{16} was explicitly designed for real-time segmentation on the Kvasir and CVC datasets, but with some loss of accuracy for the sake of speed. Jha et al.~\cite{5} then created NanoNet, which is a light model with three variants (NanoNet-A, NanoNet-B, and NanoNet-C) that balance speed and accuracy. Among these, NanoNet-A has greater accuracy with more parameters, whereas NanoNet-C puts greater emphasis on speed with fewer parameters.

\subsection{Transformers networks}

Transformers were originally proposed in natural language processing (NLP) and delivered outstanding performance~\cite{17}. Transformers use multi-head self-attention (MHSA) layers for capturing long-range dependencies. Dosovitskiy et al.~\cite{18} transferred transformers to computer vision, presenting the Vision Transformer (ViT), which represents images as a sequence of patch embeddings. Although ViT achieves good results in classification, low-resolution, single-scale feature maps are hard to use in dense prediction tasks such as segmentation and object detection.

Pyramid Vision Transformer (PVT)-based models~\cite{19, 20} overcome these shortcomings using fine-grained patches (4×4 per patch) and hierarchical pyramid architecture for high-resolution feature learning in a computationally less demanding process. Trailblazing PVT, Dong et al.~\cite{21} proposed Polyp-PVT, a polyp segmentation network that augments feature extraction using a transformer encoder and multi-level feature fusion.

Some other improvements include TransUNet~\cite{transunet}, which combines ViT with U-Net to retain global context while maintaining precise localization. Swin-UNet~\cite{swinunet} and Swin-UNETR~\cite{swinunetr} utilize better spatial efficiency through shifted window attention for enabling better scalability for high-resolution medical images. For colonoscopy segmentation, FCB-Former and its extension FCB-Former+SEP~\cite{fcbformer_sep} combine convolutional and transformer blocks to improve feature representation. Lastly, NanoNet-A/C ~\cite{nanonet_ac} provide light-weight architectures optimized for deployment on edge devices with an attempt to balance efficiency with accuracy.

\section{Proposed Architecture}
\label{sec:architecture}

Existing hybrid Transformer-CNN segmentation models hardly reach the full potential of synergies between global attention and local feature extraction. We resolve this with a thoughtfully crafted Swin Transformer-CNN architecture with three innovations: (1) adaptive fusion modules to balance transformer and CNN features adaptively across different scales, (2) context-preserving skip connections that preserve spatial accuracy, and (3) a computationally efficient cross-attention bridge. The last structure illustrates enhanced performance in medical image segmentation, where current approaches fail to capture anatomical context and fine boundaries at the same time.

\subsection{Encoder: Swin Transformer Backbone}
We used a transformer-based backbone to perform multi-scale features extraction in encoding. It starts with an input RGB image $\mathbf{X}$ of size $\mathbb{R}^{H \times W \times 3}$. It conditions the image as it passes through the backbone to increasingly four stages at different resolutions. The model in every stage produces feature maps of certain spatial sizes and channel depths.

\begin{equation}
\begin{aligned}
\mathbf{F}_1 &\in \mathbb{R}^{\frac{H}{4} \times \frac{W}{4} \times C} \quad (\text{Stage 1}) \\
\mathbf{F}_2 &\in \mathbb{R}^{\frac{H}{8} \times \frac{W}{8} \times 2C} \quad (\text{Stage 2}) \\
\mathbf{F}_3 &\in \mathbb{R}^{\frac{H}{16} \times \frac{W}{16} \times 4C} \quad (\text{Stage 3}) \\
\mathbf{F}_4 &\in \mathbb{R}^{\frac{H}{32} \times \frac{W}{32} \times 8C} \quad (\text{Stage 4}) \\
\end{aligned}
\end{equation}

where $\mathbf{F}_i$ denotes level $i$ features with decreasing spatial resolutions and deeper channels. The Swin Transformer's $\textit{shifted window self-attention}$ supports efficient modeling of long-range dependencies without the loss of computational efficiency.

\subsection{Decoder: Feature Fusion and Upsampling}

The decoder generates high-resolution predictions via upsampling and feature fusion operations. Decoder block $\mathcal{D}_i$ is comprised of:

\begin{enumerate}
    \item The features are first processed by a $3×3$ convolutional layer with batch normalisation and ReLU activation.
    \begin{equation}
        \mathbf{\hat{F}}_i = \sigma(\text{BN}(\text{Conv}_{3\times3}(\mathbf{F}_i)))
    \end{equation}
    where $\sigma$ denotes the ReLU activation function.
    
    \item Bilinear up-sampling (×2) to achieve spatial resolution restore:
    \begin{equation}
        \mathbf{\tilde{F}}_i = \text{Upsample}_{\times 2}(\mathbf{\hat{F}}_i)
    \end{equation}
\end{enumerate}

The decoder increasingly integrates features through skip connections:

\begin{equation}
\begin{aligned}
\mathbf{D}_4 &= \mathcal{D}_4(\mathbf{F}_4), \\
\mathbf{D}_3 &= \mathcal{D}_3(\text{Concat}(\mathbf{D}_4, \mathbf{F}_3)), \\
\mathbf{D}_2 &= \mathcal{D}_2(\text{Concat}(\mathbf{D}_3, \mathbf{F}_2)), \\
\mathbf{D}_1 &= \mathcal{D}_1(\text{Concat}(\mathbf{D}_2, \mathbf{F}_1)).
\end{aligned}
\end{equation}

\subsection{Final Prediction Layer}
The decoder output $\mathbf{D}_1 \in \mathbb{R}^{\frac{H}{2} \times \frac{W}{2} \times 64}$ is mapped to the target mask space through a $1\times1$ convolution:
\begin{equation}
\mathbf{Y} = \text{Conv}_{1\times1}(\mathbf{D}_1), \quad \mathbf{Y} \in \mathbb{R}^{\frac{H}{2} \times \frac{W}{2} \times K},
\end{equation}

where $K$ is the number of classes. Another bilinear interpolation generates the prediction in the input resolution:

\begin{equation}
\mathbf{\hat{Y}} = \text{Upsample}(\mathbf{Y}, \text{size}=(H, W)).
\end{equation}

\begin{table*}[ht]
\centering
\small
\caption{Comparison of publicly available polyp detection/segmentation datasets.}
\begin{tabular}{|l|l|r|l|}
\hline
\textbf{Dataset Name (Year, Country)} & \textbf{Ground Truth} & \textbf{Images} & \textbf{Resolution} \\ \hline
CVC-ColonDB (2013, Spain)~\cite{bernal2013} & Binary Mask & 380 & 500×574 \\ \hline
ETIS-LaribPolypDB (2014, France)~\cite{silva2014} & Binary Mask & 196 & 1225×966 \\ \hline
CVC-ClinicDB (2015, Spain)~\cite{bernal2015} & Binary Mask & 612 & 576×768 \\ \hline
ASU-Mayo (2016, USA)~\cite{tajbakhsh2016} & Binary Mask + BBox & 18,781 & 512×512 \\ \hline
GI Lesions (2016, France)~\cite{ali2016} & Annotated File + BBox & 30 videos & 768×576 \\ \hline
EndoScene (2016)~\cite{pogorelov2017} & Binary Mask & 912 & 224×224 \\ \hline
CVC-ClinicVideoDB (2017)~\cite{bernal2017} & Binary Mask & 11,954 frames & 384×288 \\ \hline
Kvasir-SEG (2019, Norway)~\cite{jha2019} & Binary Mask + BBox & 1,000 & 320×320 \\ \hline
KvasirCapsule-SEG (2019, Norway)~\cite{smedsrud2019} & BBox & 47,238 & Varies \\ \hline
NBIPolyp-Ucdb (2019, Portugal)~\cite{reis2019} & Binary Mask & 86 & 576×720 \\ \hline
WLPolyp-UCdb (2019, Portugal)~\cite{reis2019} & Annotated File & 3,040 & 726×576 \\ \hline
KUMC (2020, Korea)~\cite{kim2020} & BBox & 4,856 & 224×224 \\ \hline
SUN (2020, Japan)~\cite{liu2020} & BBox & 49,136 & 416×416 \\ \hline
PICCOLO (2020, Spain)~\cite{jaquezi2020} & Binary Mask & 3,433 & 854×480 \\ \hline
CP-CHILD (2020, China)~\cite{liu2020child} & Annotated File & 9,500 & 256×256 \\ \hline
EDD2020 (2020, International)~\cite{ali2020} & BBox + Binary Mask & 386 videos & Varies \\ \hline
HyperKvasir (2020, Norway)~\cite{borgli2020} & Binary Mask & 10,662 & 224×224 \\ \hline
Kvasir-Capsule (2021, Norway)~\cite{hicks2021} & BBox & 47,238 & Varies \\ \hline
LD Polyp Video (2021, China)~\cite{liu2021} & BBox & 40,187 frames & 560×480 \\ \hline
SUN-SEG (2022, Japan)~\cite{fan2022} & Multiple types & 158,690 & 416×416 \\ \hline
PolypGen (2022, Multi-center)~\cite{ali2022} & Binary Mask + BBox & 6,282 & Various \\ \hline
\end{tabular}
\label{tab:polyp_datasets}
\end{table*}

\section{Implementation details}

We implemented our hybrid model in PyTorch framework and evaluated on an NVIDIA GeForce RTX 4090 GPU with 24GB VRAM. For handling variations in polyp image sizes, a multi-scale approach was employed during training. AdamW optimizer, which is appropriate for transformer-based models, was utilized with learning rate of \(1 \times 10^{-4}\) and weight decay of \(1 \times 10^{-4}\). The loss function combined binary cross-entropy (BCE) and Intersection over Union (IoU) to optimize segmentation accuracy.

Input images were downsized to \(352 \times 352\) pixels, with a mini-batch of size 8, for 100 epochs. Training took about 1 hour, and best coscusative performance was at epoch 63. In an effort to avoid overfitting, an early stopping condition was used, inspecting the Dice score on the test set after every epoch. Training was stopped if no progress had been made in the last 37 epochs, which happened at epoch 63.

During training, the data augmentation processes like random rotation, flip left-right, and flip top-bottom were utilized. Images were rescaled to \(352 \times 352\) during test time without any extra post-processing or optimization techniques. This method possessed strong performance with low computational cost.

\section{Experiments}

\subsection{Datasets}
We used our approach to the Kvasir-SEG dataset, which contains 1000 polyp images that are subclasses of the Kvasir polyp class. We divided the dataset into 900 images for training purposes and 100 images for the test set. Training was carried out once while cross-validation was applied solely on the test set within the Kvasir dataset. As can be seen from Table ~\ref{tab:model_comparison}, we achieved a test set accuracy of 0.987.

For further testing, we have tested the model on four test datasets, i.e., CVC-ClinicDB, ETIS, CVC-ColonDB, and Endotect. The test datasets are of a variety of polyp images for which generalizability estimation is possible with reliability. ETIS contains 196 images, CVC-ColonDB contains 380 images, CVC-ClinicDB contains 612 images, and Endotect contains 1000 images. Multi-dataset testing provides a certain realization about method performance in various clinical conditions. The list of all the datasets is as depicted in figure ~\ref{tab:polyp_datasets}.

\begin{figure*}[h]
    \centering
    \includegraphics[width=\textwidth]{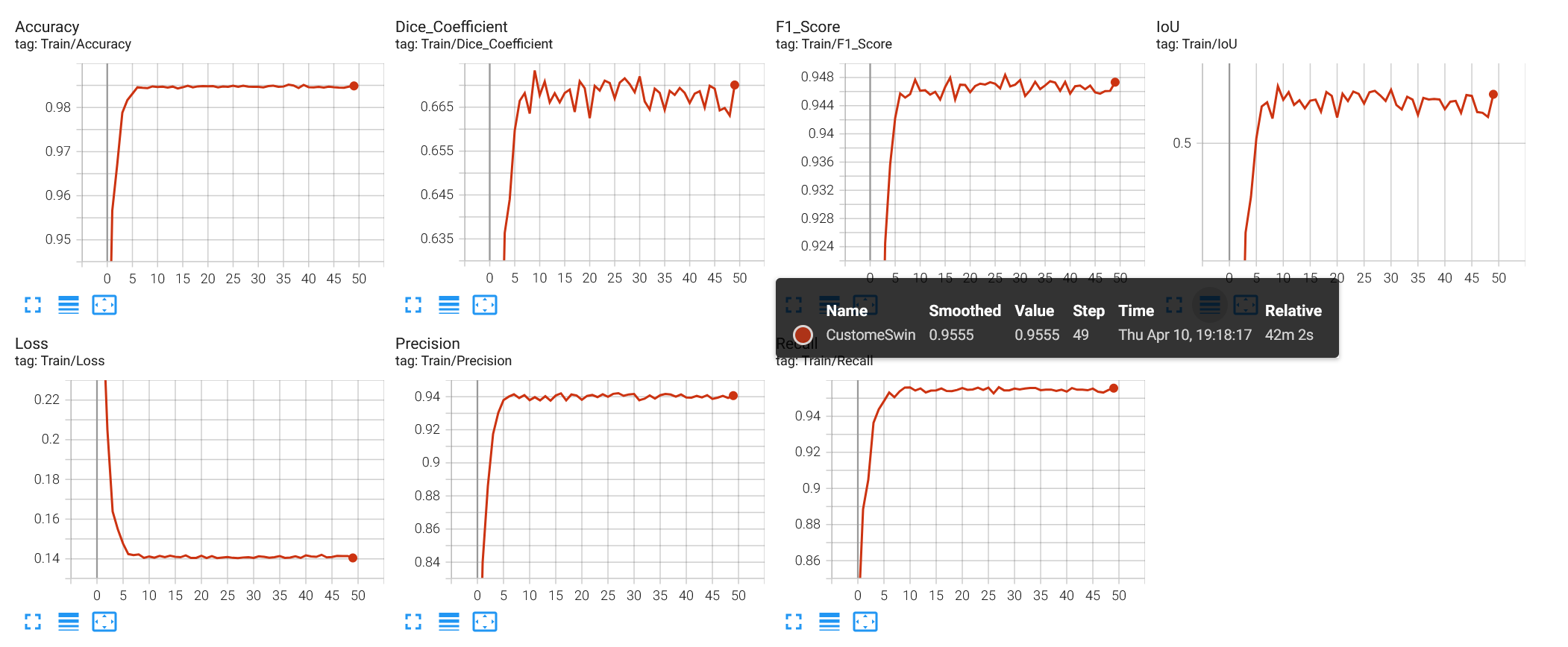} 
    \caption{Evaluation of our model in the Kvasir-SEG dataset, fifty epochs}
    \label{fig:wide_image}
\end{figure*}

Datasets:\url{https://bit.ly/polyp-datasets}

\subsection{Evaluation metrics}

For the purpose of measuring our model, our chosen metrics are: Dice Score Coefficient (DSC), mean Intersection over Union (mIoU), Precision, Recall, F2, Accuracy, and Frames-per-Second (FPS).

\begin{align}
\text{IoU} &= \frac{\text{TP}}{\text{TP} + \text{FP} + \text{FN}} \label{eq:iou} \\
\text{DSC} &= \frac{2 \times \text{TP}}{2 \times \text{TP} + \text{FP} + \text{FN}} \label{eq:dice} \\
\text{Precision} &= \frac{\text{TP}}{\text{TP} + \text{FP}} \label{eq:precision} \\
\text{Recall/Sensitivity} &= \frac{\text{TP}}{\text{TP} + \text{FN}} \label{eq:recall} \\
\text{Accuracy} &= \frac{\text{TP} + \text{TN}}{\text{TP} + \text{TN} + \text{FP} + \text{FN}} \label{eq:accuracy} \\
\text{F1-score} &= \frac{2 \times \text{Precision} \times \text{Recall}}{\text{Precision} + \text{Recall}} \label{eq:f1}  
\end{align}

\section{Results}

We evaluated our model and compared its performance against recent state-of-the-art (SOTA) polyp segmentation models. The evaluation metrics, as shown in table ~\ref{tab:model_comparison}, were used for benchmarking. On the Kvasir-SEG dataset, our method achieved a recall score of 0.9555, which is 1.76 \% higher than the existing real-time SOTA method DUCK-Net. Similarly, the accuracy reached 0.9849, reflecting a little improvement in the existing DUCK-Net.

\begin{table*}[t]
\caption{Performance Comparison of Polyp Segmentation Models in Kavasir-SEG dataset}
\label{tab:model_comparison}
\centering
\small
\begin{tabular}{@{}lccccccc@{}}
\toprule
\textbf{Model} & \textbf{F1 Score} & \textbf{mDice} & \textbf{mIoU} & \textbf{Precision} & \textbf{Recall} & \textbf{Accuracy} \\
\midrule
U-Net\textsuperscript{\dag} ~\cite{13} & 0.8655 & -- & 0.7629 & 0.8593 & 0.8718 & - &  \\
ResUNet ~\cite{zhang2018road} & 0.7878 & -- & 0.7778 & -- & -- & \\
ResUNet++ ~\cite{7} & 0.8133 & -- & 0.7927 & 0.7064 & 0.8774 & - &\\
Li-SegPNet ~\cite{li2022segpnet} & 0.9058 & -- & 0.8800 & 0.9424 & 0.9254 & - & \\
PraNet ~\cite{fan2020pranet} & 0.9094 & -- & 0.8339 & 0.9599 & 0.8640 & - & \\
ColonFormer ~\cite{colonformer2022} & -- & 0.927 & 0.877 & -- & -- & - & \\
DUCK-Net ~\cite{ducknet2023} & 0.9502 & -- & 0.9051 & 0.9628 & 0.9379 & 0.9842 & \\
Hybrid(Our) & 0.9499 & - & - & 0.9422 & \textbf{0.955} & \textbf{0.9849} &\\
\bottomrule
\end{tabular}
\end{table*}

\section{Conclusion}

In this paper, we present high accurate image polyp segmentation, called Hybrid(Transformer + CNN), which incorporates a vision transformer backbone for efficient feature extraction with CNN skip connection layers. Experimental results on various endoscopy datasets demonstrate that our model achieves state-of-the-art performance across key metrics, including DSC, IoU, precision, recall, F2-score, and, crucially, FPS with resonalbale model parameters and inference speed.

We believe hybrid architecture offers significant potential for detecting pathological and abnormal tissues within the colon lining. One of its key advantages is the ability to identify flat polyps in challenging regions like variation in size, shape, endoscopy types, lighting, imaging protocols of the colon and detect small lesions that might be overlooked during standard endoscopy. Besides this, it can also help differentiate residual tissue after polyp removal during colonoscopy, ensuring total removal and reducing the recurrence risk.

We hope our work inspires other researchers to tackle real-time polyp segmentation tasks using hybrid-based networks. Beyond endoscopy, we envision hybrid architecture being applied in other medical fields. For example, it can assist with early diagnosis of polyp. By enhancing outcomes in cases that are under subjective clinical judgments, our technique might maximize patient therapy.

We believe our proposed method has broad implications and can contribute to advancing medical imaging and intervention techniques across multiple specialties.

{\small
\bibliographystyle{ieee_fullname}
\bibliography{egbib}
\nocite{*}
}

\end{document}